\documentclass[aps,pra,twocolumn,groupedaddress,amsmath,amssymb]{revtex4}
\usepackage{graphicx}
\begin{document}

\title{Quantum Teleportation with Atoms Trapped in Cavities}
\author{Jaeyoon Cho}
\author{Hai-Woong Lee}
\affiliation{Department of Physics, Korea Advanced Institute of Science and Technology, Daejeon 305-701, Korea}
\date{\today}
\begin{abstract}
We propose a scheme to implement the quantum teleportation protocol with single atoms trapped in cavities.
The scheme is based on the adiabatic passage and the polarization measurement.
We show that it is possible to teleport the internal state of an atom trapped in a cavity to an atom trapped in another cavity with the success probability of $1/2$ and the fidelity of 1.
The scheme is resistant to a number of considerable imperfections such as the violation of the Lamb-Dicke condition, weak atom-cavity coupling, spontaneous emission, and detection inefficiency.
\end{abstract}
\pacs{pacs}
\maketitle

\newcommand{\bra}[1]{\left<#1\right|}
\newcommand{\ket}[1]{\left|#1\right>}
\newcommand{\abs}[1]{\left|#1\right|}

Recent advances in cavity quantum electrodynamics made it possible to trap and manipulate single atoms in high-$Q$ cavities
\cite{ye99,mckeever03,maunz04} and thereby to achieve various modes of quantum information processing with single trapped atoms.
There have been numerous proposals to use single trapped atoms for quantum information processing, such as for entanglement generation \cite{hong02,feng03}, quantum computation \cite{pellizzari95}, quantum communication \cite{pellizzari97}, and for quantum teleportation \cite{bose99}, the topic of our present investigation.

Since the first proposal \cite{bennett93}, quantum teleportation of polarization states of single photons and coherent states of light has been demonstrated experimentally \cite{furusawa98}.
Experimental demonstration of teleportation of atomic states, however, is yet to be realized.
In earlier proposals of quantum teleportation of atomic states \cite{zheng99}, qubits were internal states of single flying atoms.
From the viewpoint of quantum information processing, however, it would be ideal to have atoms as stationary qubits used only for storage of information and leave communication to photons.
The above mentioned advances in cavity quantum electrodynamics techniques of trapping and manipulating atoms open ways for such a scheme.

Bose \textit{et al.} \cite{bose99} in 1999 proposed an attractive scheme to teleport an atomic state in which atoms trapped in cavities play the role of stationary qubits and photons leaking out from the cavities are used for Bell-state measurements. As this scheme is relatively easy to implement experimentally and has several practical advantages over the schemes proposed earlier, as described in Ref. \cite{bose99}, experimental demonstration of the atomic-state teleportation seems feasible within the near future, provided that a few difficulties still facing the scheme, such as the state-dependent success probability and fidelity and the necessity to count photons, are resolved.

In this letter, we propose a scheme that overcomes most of the difficulties of the scheme proposed by Bose \textit{et al}. The basic idea of our proposed scheme is similar to that of their scheme, so that it inherits the simplicity and easiness to implement experimentally. As in their scheme, we use photons leaking out from the cavities for Bell-state measurements. The main difference, however, lies in the use of photon polarization states, rather than the one-photon and vacuum states as in Bose \textit{et al.}, to represent quantum bits. On the one hand, this difference leads necessarily to a somewhat more complicated atomic level structure to be exploited for teleportation. On the other hand, as will be described in detail below, it opens a way for significant improvements that allow to achieve an ideal success probability of $\frac{1}{2}$ (1 if nonlinear interactions are exploited in Bell-state measurements) and an ideal fidelity of 1. Moreover, our scheme does not require distinguishing between one and two photons.

\begin{figure}[b]
\includegraphics[width=0.7\columnwidth]{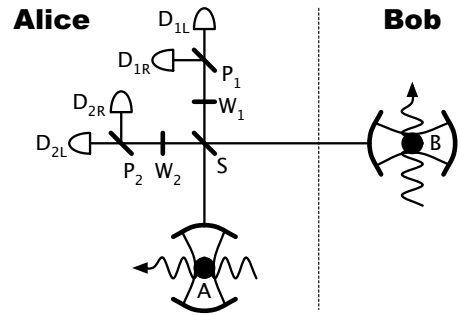}
\caption{\label{fig:scheme}Experimental scheme to teleport the internal state of atom $A$ to atom $B$. $S$ is a beam splitter, $W_{1}$ and $W_{2}$ are quarter wave plates, $P_{1}$ and $P_{2}$ are polarization beam splitters, and $D_{1L}$, $D_{2L}$, $D_{1R}$, and $D_{2R}$ are photodetectors. Each winding arrow represents the classical driving field}
\end{figure}

The schematic representation of our scheme is shown in Fig.~\ref{fig:scheme}.
The atom A is trapped in Alice's cavity, and the atom B in Bob's cavity.
Each atom is driven adiabatically by a classical coherent field.
Alice maps the unknown internal state of her atom into the two-mode state of her cavity through adiabatic passage \cite{kuhn02}, while Bob generates a maximally entangled state of the internal state of his atom and the two-mode state of his cavity through adiabatic passage.
During both the adiabatic passage processes, with the probability of 1, each cavity should emit one photon with two possible polarization degrees of freedom in which the quantum information is encoded.
Two photons leaking out from both cavities interfere at the 50-50 beam splitter
$S$ at Alice's site.
The beam splitter $S$, two quarter wave plates $W_{1}$ and $W_{2}$, two
polarization beam splitters $P_{1}$ and $P_{2}$, and four detectors $D_{1L}$,
$D_{1R}$, $D_{2L}$, and $D_{2R}$ constitute a measurement device for
discriminating between the Bell states of the two single-photon polarization
qubits.

\begin{figure}[b]
\includegraphics[width=0.7\columnwidth]{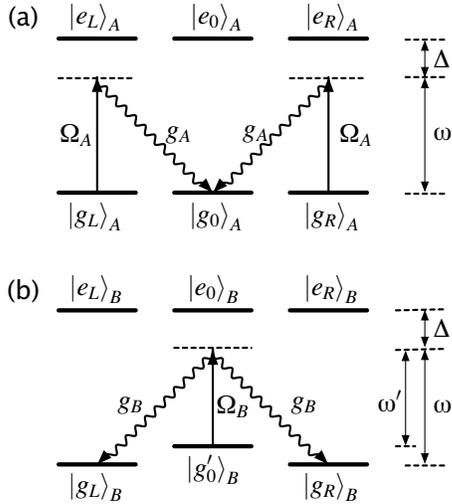}
\caption{\label{fig:level}The involved atomic levels and transitions for Alice (a) and Bob (b). Alice's qubit is encoded in the two Zeeman sublevels $\ket{g_{L}}_{A}$ and $\ket{g_{R}}_{A}$, and Bob's qubit in the same way. Each straight arrow represents the transition driven by the $\pi$-polarized classical coherent field and each winding arrow represents the transition due to the atom-cavity coupling. Each transition of $\ket{e_{L}}_{A}\rightarrow\ket{g_{0}}_{A}$ and $\ket{e_{0}}_{B}\rightarrow\ket{g_{R}}_{B}$ ($\ket{e_{R}}_{A}\rightarrow\ket{g_{0}}_{A}$ and $\ket{e_{0}}_{B}\rightarrow\ket{g_{L}}_{B}$) is coupled to the left-circularly (right-circularly) polarized mode of the cavity. The transition $\ket{g_{0}}_{A}\leftrightarrow\ket{e_{0}}_{A}$ is electric dipole forbidden.}
\end{figure}

The involved atomic levels and transitions are depicted in Fig.~\ref{fig:level}.
For the operation, both Alice and Bob exploit two $F=1$ hyperfine levels, whereas Bob exploits one additional hyperfine level.
A qubit is encoded in two Zeeman sublevels of the $F=1$ ground hyperfine level.
To express the state of the atom-cavity system, we use the notation
\begin{equation}
\ket{\Psi(t)}_{i}=\ket{x}_{i}\ket{n_{L},n_{R}}_{i},
\end{equation}
where $i=A,B$ denotes Alice or Bob, $x$ the atomic state,
and $n_{L,R}$ the number of left- or right-circularly polarized photons in
Alice's or Bob's cavity.
$\Omega_{i}(t)$ and $g_{i}$ represent the time-dependent Rabi frequency of the classical field and the atom-cavity coupling rate (assumed to be the same for both the transitions), respectively, with $i=A,B$ for Alice or Bob.
Both classical and quantum fields are detuned from the atomic resonance by an amount of $\Delta$.
For the moment, we assume that $g_{i}$ remains constant during the operation.
The assumption is valid in the Lamb-Dicke limit.
The effect of time-varying $g_{i}$ will be discussed later.

Initially, Alice's system is prepared in the following state:
\begin{equation}
\ket{\Psi(0)}_{A}=(\alpha\ket{g_{L}}+\beta\ket{g_{R}})_{A}\ket{0,0}_{A}\label{eq:alice},
\end{equation}
where $\alpha$ and $\beta$ are unknown.
If the variation of $\Omega_{A}(t)$ is sufficiently slow, only the four transitions are involved as depicted in Fig.~\ref{fig:level}(a): $\ket{g_{m}}_{A}\rightarrow\ket{e_{m}}_{A}$ ($m=L,R$) driven by the $\pi$-polarized classical field and $\ket{e_{L}}_{A}\rightarrow\ket{g_{0}}_{A}$ ($\ket{e_{R}}_{A}\rightarrow\ket{g_{0}}_{A}$) coupled to the left-circularly (right-circularly) polarized mode of the cavity.
The transition between $\ket{g_{0}}_{A}$ and $\ket{e_{0}}_{A}$ is electric dipole forbidden.
Consequently, in the rotating frame, the Hamiltonian of the total system can be written as
\begin{eqnarray}
H_{A}&=&-(\Delta+i\gamma_A/2)(\ket{e_L}\bra{e_L}+\ket{e_R}\bra{e_R})_A\nonumber\\
& &+[\Omega_{A}(t)(\ket{e_{L}}\bra{g_{L}}+\ket{e_{R}}\bra{g_{R}})_{A}\nonumber\\
& &+g_{A}(a_{L}^{A}\ket{e_{L}}\bra{g_{0}}+a_{R}^{A}\ket{e_{R}}\bra{g_{0}})_{A}+h.c.],
\end{eqnarray}
where $\gamma_A$ and $a_{L,R}^{A}$ denote the atomic spontaneous emission rate and the annihilation operator for the corresponding polarized mode of the cavity, respectively.
The dark space is spanned by the two eigenstates $\ket{D_{1}(t)}_{A}=\cos\theta_{A}(t)\ket{g_{L}}_{A}\ket{0,0}_{A}-\sin\theta_{A}(t)\ket{g_{0}}_{A}\ket{1,0}_{A}$ and $\ket{D_{2}(t)}_{A}=\cos\theta_{A}(t)\ket{g_{R}}_{A}\ket{0,0}_{A}-\sin\theta_{A}(t)\ket{g_{0}}_{A}\ket{0,1}_{A}$, where $\theta_{A}(t)$ is given by $\cos\theta_{A}(t)=\frac{g_{A}}{\sqrt{\abs{g_{A}}^{2}+\abs{\Omega_{A}}^{2}}}$ and $\sin\theta_{A}(t)=\frac{\Omega_{A}(t)}{\sqrt{\abs{g_{A}}^{2}+\abs{\Omega_{A}}^{2}}}$.
In the adiabatic limit, the initial state~(\ref{eq:alice}) evolves in the dark space into the following state:
\begin{eqnarray}
\ket{\Psi(t)}_{A}&=&\alpha\ket{D_{1}(t)}_{A}+\beta\ket{D_{2}(t)}_{A}\nonumber\\
&=&\cos\theta_{A}(t)(\alpha\ket{g_{L}}+\beta\ket{g_{R}})_{A}\ket{0,0}_{A}\nonumber\\
& &-\sin\theta_{A}(t)\ket{g_{0}}_{A}(\alpha\ket{1,0}+\beta\ket{0,1})_{A}.
\end{eqnarray}
Alice, thus, can map her atomic state $(\alpha\ket{g_{L}}+\beta\ket{g_{R}})_{A}$ into her cavity mode state $(\alpha\ket{1,0}+\beta\ket{0,1})_{A}$ by simply increasing $\sin\theta_{A}(t)$ adiabatically.

For Bob, the atom is initially prepared in the state $\ket{g_{0}'}_{B}\ket{0,0}_{B}$.
The process for Bob is similar to that for Alice.
With $\Omega_{B}$ varied adiabatically, only the three transitions are involved as depicted in Fig.~\ref{fig:level}(b): $\ket{g_{0}'}_{B}\rightarrow\ket{e_{0}}_{B}$ driven by the $\pi$-polarized classical field and $\ket{e_{0}}_{B}\rightarrow\ket{g_{L}}_{B}$ ($\ket{e_{0}}_{B}\rightarrow\ket{g_{R}}_{B}$) coupled to the right-circularly (left-circularly) polarized mode of the cavity.
Consequently, in the rotating frame, the Hamiltonian of the total system can be written as
\begin{eqnarray}
H_{B}&=&-(\Delta+i\gamma_B/2)(\ket{e_0}\bra{e_0})_B\nonumber\\
& &+[\Omega_{B}(t)(\ket{e_{0}}\bra{g_{0}'})_{B}\nonumber\\
& &+g_{B}(a_{R}^{B}\ket{e_{0}}\bra{g_{L}}+a_{L}^{B}\ket{e_{0}}\bra{g_{R}})_{B}+h.c.],
\end{eqnarray}
where $\gamma_B$ and $a_{L,R}^{B}$ denote the atomic spontaneous emission rate and the annihilation operator for the corresponding polarized mode of the cavity, respectively.
In the adiabatic limit, the initial state evolves into the following dark state:
\begin{eqnarray}
\ket{\Psi(t)}_{B}&=&\cos\theta_{B}(t)\ket{g_{0}'}_{B}\ket{0,0}_{B}-\nonumber\\
& &\sin\theta_{B}(t)\frac{\ket{g_{L}}_{B}\ket{0,1}_{B}+\ket{g_{R}}_{B}\ket{1,0}_{B}}{\sqrt{2}},
\end{eqnarray}
where $\theta_{B}(t)$ is given by $\cos\theta_{B}(t)=\frac{\sqrt{2}g_{B}}{\sqrt{2\abs{g_{B}}^{2}+\abs{\Omega_{B}}^{2}}}$ and $\sin\theta_{B}(t)=\frac{\Omega_{B}(t)}{\sqrt{2\abs{g_{B}}^{2}+\abs{\Omega_{B}}^{2}}}$. 
Bob also increase $\sin\theta_{B}(t)$ adiabatically to generate a maximally entangled state $\frac{(\ket{g_{L}}\ket{0,1}+\ket{g_{R}}\ket{1,0})_{B}}{\sqrt{2}}$.

As $\sin\theta_{A}(t)$ and $\sin\theta_{B}(t)$ are increased, each cavity emits one photon at some instant.
To illustrate the basic idea of our scheme, let us first assume that both the photons reach simultaneously at the beam splitter $S$.
Expressing the polarizations of each photon as $\ket{L}_{i}$ and $\ket{R}_{i}$ respectively, with $i=A,B$ for Alice or Bob, the total state can be written as
\begin{equation}
\ket{\Psi'}=\frac{1}{\sqrt{2}}\ket{g_{0}}_{A}(\alpha\ket{L}+\beta\ket{R})_{A}(\ket{R}\ket{g_{L}}+\ket{L}\ket{g_{R}})_{B}.
\end{equation}
Now it is clear that a Bell-state measurement with two single-photon polarization qubits followed by the corresponding unitary operation to Bob's atom completes the quantum teleportation.
As we consider only linear optical elements, the success probability of such a Bell measurement is limited up to $1/2$ \cite{lutkenhaus99}.
If nonlinear interaction between two photons is available, the success probability could be as high as 1 \cite{nonlinear}.
In our setup of Fig.~\ref{fig:scheme}, the Bell-state measurement succeeds only when the two photons are found to be oppositely polarized at two detectors.
From simple calculations, it is found that when $D_{1L}$ and $D_{1R}$ click or $D_{2L}$ and $D_{2R}$ click, the state of Bob's atom collapses into the state $\alpha\ket{g_{L}}_{B}+\beta\ket{g_{R}}_{B}$, whereas when $D_{1L}$ and $D_{2R}$ click or $D_{2L}$ and $D_{1R}$ click, into the state $\alpha\ket{g_{L}}_{B}-\beta\ket{g_{R}}_{B}$.
For the latter case, Bob applies an appropriate local unitary operation to his atom to transform the state into the former one.

We return to the actual situation in which each photon leaks out from the cavity in the form of a single-photon pulse due to the random nature of the emission.
In the adiabatic limit, the pulse shape can be calculated as \cite{duan03}
\begin{equation}
f_{i}(t)=\sqrt{\kappa_{i}}\sin\theta_{i}(t)\exp\left(-\frac{\kappa_{i}}{2}\int_{0}^{t}\sin^{2}\theta_{i}(\tau)d\tau\right)\label{eq:pulse},
\end{equation}
where $\kappa_{i}$ denotes the cavity decay rate for Alice ($i=A$) or Bob ($i=B$).
The two photons interfere maximally when the two pulse shapes overlap completely at the beam splitter.
Therefore, in order to get the maximum fidelity of 1, $\Omega_{i}(t)$ should be adjusted to satisfy $\sin\theta_{A}(t)=\sin\theta_{B}(t)$, where we have assumed that two distances between the cavities and the beam splitter are the same.
We do not have to adjust the exact pulse shape.
From a simple algebra, we obtain the following condition:
\begin{equation}
\Omega_{B}(t)=\sqrt{2}\abs{\frac{g_{B}}{g_{A}}}\Omega_{A}(t).\label{eq:cond}
\end{equation}

We now compare our scheme with that of Bose \textit{et al.} \cite{bose99} in detail. The success of the teleportation process in the scheme by Bose \textit{et al.} requires first a successful preparation of the atom-field states for both Alice and Bob, which in turn requires no decay of photons from both Alice's and Bob's cavities being recorded. The detection that follows the successful preparation can lead to successful teleportation if, during the detection period, one and only one photon is detected by one of Alice's detectors, whereas there is a possibility that no photon is emitted or two photons are emitted from the cavities. Even if one photon is detected, there still exists a possibility that teleportation fails. The total success probability is then given by the product of the success probabilities at each different stage, and thus depends on the initial state and system parameters in a complicated manner. In contrast, our scheme does not require a separate preparation stage. Detection begins at the instant at which adiabatic pulses are turned on. One only needs to apply the adiabatic pulses to the state~(\ref{eq:alice}) at Alice's site and to the state $\ket{g_{0}'}_{B}\ket{0,0}_{B}$ at Bob's site, and simply wait for two photons to be detected by the detectors. Detection of two photons at {\em any arbitrary instant} $t\ge 0$ guarantees the success of the teleportation process. Since our scheme is designed so that the number of photons leaking out from the cavities is fixed as two (a similar design has been used in the recent entanglement generation scheme \cite{feng03}), the detection of two photons is bound to occur, resulting in an ideal success probability of $\frac{1}{2}$ (assuming linear optical methods for Bell-state measurements) and an ideal fidelity of 1, provided one has perfect detectors. Even if the detection inefficiency is considered, the fidelity is not affected (when we ignore dark counts), although the success probability is lowered. Another important practical advantage of our scheme is that photon counting is not required. Detection of photons at two different detectors is sufficient to guarantee a faithful teleportation (when we ignore dark counts).

Our scheme has additional favorable features that it works fine outside the Lamb-Dicke regime and in the weak atom-cavity coupling regime.
In order to show this, we present below results of our numerical simulations performed using the method described in Ref. \cite{duan03}. In our numerical simulations, we allow $g_{i}$ to vary in time.
The change of $g_{i}$ is accompanied with the change of the output pulse shape $f_{i}(t)$.
From a straightforward calculation, we find out that the fidelity $F$ of the
resulting state of Bob's atom with respect to the initial state of Alice's
atom is determined by the overlap of the two pulse shapes as the following:
\begin{equation}
F=\sqrt{\abs{\alpha}^{4}+\abs{\beta}^{4}+\frac{1}{P}\left|\alpha\beta\!
\int_{0}^{\infty}\!dt\, f_{A}^{*}(t)f_{B}(t)\right|^2},
\label{eq:fidel}
\end{equation}
where $P=\frac{1}{2}\int_0^\infty dt_1\int_0^\infty
dt_2\abs{f_A(t_1)}^2\abs{f_B(t_2)}^2$ is the success probability.
To give a typical value of $F$, we consider a more specific situation in which
each atom is placed inside a far-off resonance trapping (FORT) potential
\cite{ye99,mckeever03,maunz04}.
A FORT beam generates multiple potential wells along the cavity axis, and each
potential well gives rise to a different value of the atom-cavity coupling rate.
For simplicity, we assume that each atom is trapped in the same potential well
and subjected to a simple harmonic oscillation.
Noting that the radial motion of the atom is much slower than the axial one \cite{ye99}, we approximate the time-dependence of $g_{i}$ as
$g_{i}(t)=g_{i0}\cos\left[\Delta_g\sin(\omega_{z}t+\phi_{i})\right]$,
where $g_{i0}$, $\Delta_g$, and $w_{z}$ denote the maximum value of
$g_{i}(t)$, the fluctuation of $g_i(t)$, and
the oscillation frequency of the atom along the cavity axis, respectively, and
an arbitrary phase $\phi_i$ is introduced to consider the random
atomic position. 
The classical field that we consider is of the Gaussian form:
$\Omega_{i}(t)=\Omega_{i0}\exp\left[-\left(\frac{t-t_{c}}{\Delta_{t}}
\right)^{2}\right]$.
The state to be teleported is chosen as $\ket{\Psi(0)}_A=\frac{\ket{g_L}_A+\ket{g_R}_A}{\sqrt{2}}$.

\begin{figure}
\includegraphics[width=0.7\columnwidth]{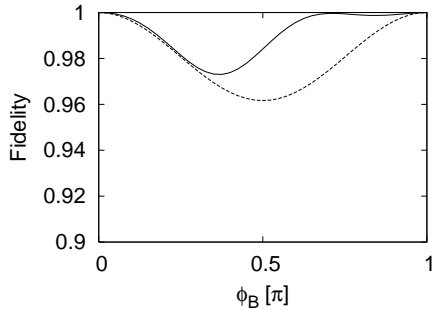}
\caption{\label{fig:fidel}Fidelity with respect to $\phi_B$ outside the Lamb-Dicke regime for the case of relatively strong atom-cavity coupling (solid curve) and weak atom-cavity coupling (dotted curve).}
\end{figure}

As a first example, we take the parameters of a typical cavity QED experiment \cite{ye99}, in which $\kappa_i/2\pi\equiv\kappa/2\pi=4~\textrm{MHz}$, $\gamma_i=\kappa$, $g_{i0}=8\kappa$, $\Delta_g=\pi/3$, and $\omega_z=0.05\kappa$.
We choose other parameters as $\Omega_{A0}=\Omega_{B0}/\sqrt{2}=g_{i0}$ (in accordance with the condition (\ref{eq:cond})), $\Delta=0$, $t_c=0.6~\mathrm{\mu s}$, and $\Delta_t=0.2~\mathrm{\mu s}$.
In Fig.~\ref{fig:fidel} we show with a solid curve the fidelity $F$ we computed with respect to $\phi_B$ while fixing $\phi_A$ as zero.
We see that the fidelity is very high for all values of $\phi_B$.
We have also computed the success probability $P$ and found it to be always very close to the ideal value ($P>0.49$). 
Our numerical simulation also indicates that emission of photons from the cavities is completed within the time of $\sim0.8~\mathrm{\mu s}$.
Since this operation time is of the same order as the period of the atomic motion ($\sim1/\omega_z$), the Lamb-Dicke condition is not satisfied and yet both the fidelity and the success probability remain high.

For our second example, we choose $g_{i0}=\gamma_i=\kappa$ ($\kappa/2\pi=4~\mathrm{MHz}$), in order to show that our scheme can work also for the case of weak atom-cavity coupling.
In this case, however, we need to have a large detuning and a long operation time to ensure that the adiabatic condition does not break down and spontaneous emission has negligible effects upon the dynamics of the system.
We therefore choose $\Delta=40\kappa$, $\Delta_t=40~\mathrm{\mu s}$, and $t_c=120~\mathrm{\mu s}$.
We also choose $\Omega_{A0}=\Omega_{B0}/\sqrt{2}=6g_{i0}$ and all other parameters to be the same as given above.
The fidelity computed is shown with a dotted curve in Fig.~\ref{fig:fidel}.
We see that the fidelity still remains very high.
The success probability in this case is also found to be very close to the ideal value ($P>0.49$). 
In this case the operation time, which is found to be $\sim70~\mathrm{\mu s}$ according to our numerical simulation, is much longer than the period of the atomic motion, and thus the Lamb-Dicke condition is violated as in the first example.
Noting that the initial state chosen gives rise to the lowest fidelity (the fidelity given by Eq.~(\ref{eq:fidel}) is the lowest when $\abs{\alpha}=\abs{\beta}=1/\sqrt{2}$), we conclude that our scheme allows the quantum teleportation to be performed with a very high fidelity and a nearly ideal success probability even outside the Lamb-Dicke regime and in the weak atom-cavity coupling regime.

Finally, we suggest ${}^{87}$Rb as a good candidate for experimental realization of our scheme.
As applied to our scheme, the states ($5^{2}\!S_{1/2}$,$F=1$) and ($5^{2}\!P_{3/2}$,$F=1$) of ${}^{87}$Rb correspond to the $F=1$ ground and excited hyperfine levels we considered, respectively, and the state ($5^{2}\!S_{1/2}$,$F=2$,$m=0$) corresponds to $\ket{g_{0}'}_{B}$.
Since the atomic properties of ${}^{87}$Rb are similar to those of ${}^{85}$Rb considered in Ref.~\cite{maunz04}, we expect that ${}^{87}$Rb can also be trapped and manipulated at the single-atom level with far-off resonance trapping within the current technology.

\end{document}